\documentclass[final]{aipproc}

\layoutstyle{8x11single}

\SetInternalRegister\hbadness{8000} % pseudo latin isn't breaking very well :-)

\newcommand\doingARLO[2][]{%
  \ifx\mmref\undefined #1\else #2\fi
}

\begin{document}

\title 
      [SDAMS: SPOrt Data Archiving and Management System]
      {SDAMS: SPOrt Data Archiving and Management System}


\author{Luciano Nicastro}{
  address={IFCAI-CNR, Via U. La Malfa 153, 90146 Palermo, Italy},
  email={nicastro@ifcai.pa.cnr.it} %,
%  thanks={This work was commissioned by the AIP}
}

\iftrue
\author{Giorgio Calderone}{
  address={IFCAI-CNR, Via U. La Malfa 153, 90146 Palermo, Italy},
  email={giorgio@ifcai.pa.cnr.it},
}
\fi

% \copyrightholder{Acoustical Scociety of America}
\copyrightyear  {2001}

\begin{abstract}
SDAMS is the ensemble of database + software packages aimed to the archiving,
quick-look analysis, off-line analysis, network accessibility and plotting of
 the SPOrt produced data. Many of the aspects related to data archiving,
 analysis and distribution are common to almost all the astronomical
 experiments. SDAMS ambition is to face and solve problems like accessibility
 and portability of the data on any hardware/software platform in a way as
 simpler as possible, though effective. The system is conceived in a way
 to be used either by the scientific community interested in background
 radiation studies or by a wider public with low or null knowledge of the
 subject. The user authentication system allows us to apply different levels
 of access, analysis and data retrieving. SDAMS will be accessible through
 any Web browser though the most efficient way to use it is by writing
 simple programs. Graphics and images useful for outreach purposes will be
 produced and put on the Web on a regular basis.
\end{abstract}

\date{\today}

\maketitle

\section{Introduction}

SPOrt (\cite{carretti:2002}) on ISS is an experiment which will
produce a limited amount of data with respect to other space experiments.
The foreseen total (i.e. housekeepings and scientific data) bit rate
is $\sim 2.5$ kbit s$^{-1}$. This means that in 3 years of operation time
it will collect $\sim 30$ Gbytes of data.
In addition SPOrt has not pointing capabilities
and it is expected to work in a stable and semi-automatic way.
As a consequence the output data stream will be simply identified by
{\em time} and {\em orbital parameters} of the Space Station.
The data management system (i.e. quick look, archive and scientific analysis)
can be build to be both simple and highly automatic.

The SDAMS system has two fundamental parts: {\em database server} and
{\em application server}.
The former performs the data storing and allows a simple (low level) fast
access to them.
To this aim we chose MySQL$^{\rm TM}$, a very easy to use data management
system capable to deal with a relatively large amount of data.
On the Web there are several examples of intensive usage of MySQL with
application tools and there are extensions to use it within several
programming languages (e.g. PHP).
It has also the advantage to be an Open Source package, that means one can
modify the source code to match any particular need and no license is
required to use it.
This is crucial to allow a standard (system) to be developed.

The SDAMS application server is a multithreaded server which uses a
simple C-written communication protocol to interact with its clients.
A task-oriented module library allows us to easily add further capabilities
to those so far foreseen.
The fundamental task of this component is to accept client connections
(through socket) and invoke the appropriate modules  to perform the
required tasks.

News about SDAMS will be posted on \url{http://sport.tesre.bo.cnr.it}.

\section{How to use it}

There are several ways to use SDAMS, the simplest being through a Web
interface.
Such kind of interface allows users to setup a simple query in order
to request data retrieval and visualization.
This is done, for example, selecting a time or sky coordinates range.
It will also be possible to view the sky portion seen by SPOrt
(pointing at the zenith of the ISS) in any given time interval.
In addition to Sun, Moon and some Planets, also "interesting" strong radio
sources will be highlighted in the sky maps.
Access to the data will be granted only to registered users.
However, this is the less flexible way to use SDAMS as it only offers a
limited set of choices; it will be likely used to perform data "quick-look"
or by users not directly involved in the SPOrt project or by
the general public.
The communication protocol is so easy that it is possible to connect to SDAMS
through a telnet session and typing in the commands.
The output will be ASCII formatted.
The most flexible way to access SDAMS is instead through a dedicated,
user written (in any language) program.
The only requirements are

\begin{itemize}
 \item capability to open TCP/IP sockets,
 \item capability to read ASCII and binary files.
\end{itemize}

In this case the output can be formatted as a FITS binary (or ASCII) table.
For these "high-level" users, a reference manual will be available
reporting characteristics of the protocol and a full description of
commands and message codes.
Once again we note how SDAMS does not put any restriction to the
hardware/software platform of the client.

\section{How it works}

Figures 1 and 2 show how SDAMS works:
for each client successfully connected to the socket a new thread is created;
 it is a sort of sub-process running on the server.
 It waits for queries from the client and executes them independently of what
 the other clients are doing.
 To each query, written by using high level commands, corresponds a spawned
 process which passes a reformatted low-level query to the database server.
 The output data are processed either by C or IDL$^{\rm TM}$ routines
(depending on the request) and then sent to the client in the requested
 format.
 This process is repeated until a "disconnect" request is received.
 The "language" used by the client to put its query is very simple, being
 an ASCII string containing the name of the procedure to use and its
 parameters.
 For example, to select the data for the Stokes parameter Q in the time
 interval (MJD) 52810--52813, in the 32 GHz frequency channel, averaged over
 30 s, and perform a polynomial fit (third order) with automatic graphical
 output, the command is:

\medskip
\begin{tt}
  sel, par='q', dt=[52810.0,52813.0], fr=32, av=30, fit='poly 3'
\end{tt}
\medskip
\\
By default the graphics goes into a PostScript file, but other formats can be
chosen.

It is worth to note how, in spite the application server talks to the
 database server using the SQL (Structured Query Language) syntax,
 it is not necessary to write
 the queries using this low-level format (but can be done starting the
 query with {\tt SQL}).
 A dedicated module, the {\em SQL wrapper}, takes care of translating
 the input command string (after a validity check).
 When the client queries require complex operations (e.g. correlation
 analysis or sky-map production), the application server first extracts
 the data into a
 local directory then the appropriate (IDL) routine (one or more) is
 invoked for the
 analysis and finally the output is sent to the client (e.g. in FITS format).
 The clients do not need an IDL license as the application software only runs
 on the server.
Figure 3 shows an example of sky coverage map extracted using the command
"{\tt map, /coverage}". It assumes only about 20 days are elapsed since
 the start of the mission.
Adding a new feature to the system is extremely easy: it is enough to add
 the IDL routine (accessing the data using the SDAMS standards) to the
 library and adding its name to the list of allowed commands.

\begin{figure}
  \includegraphics[height=.8\textheight,keepaspectratio]{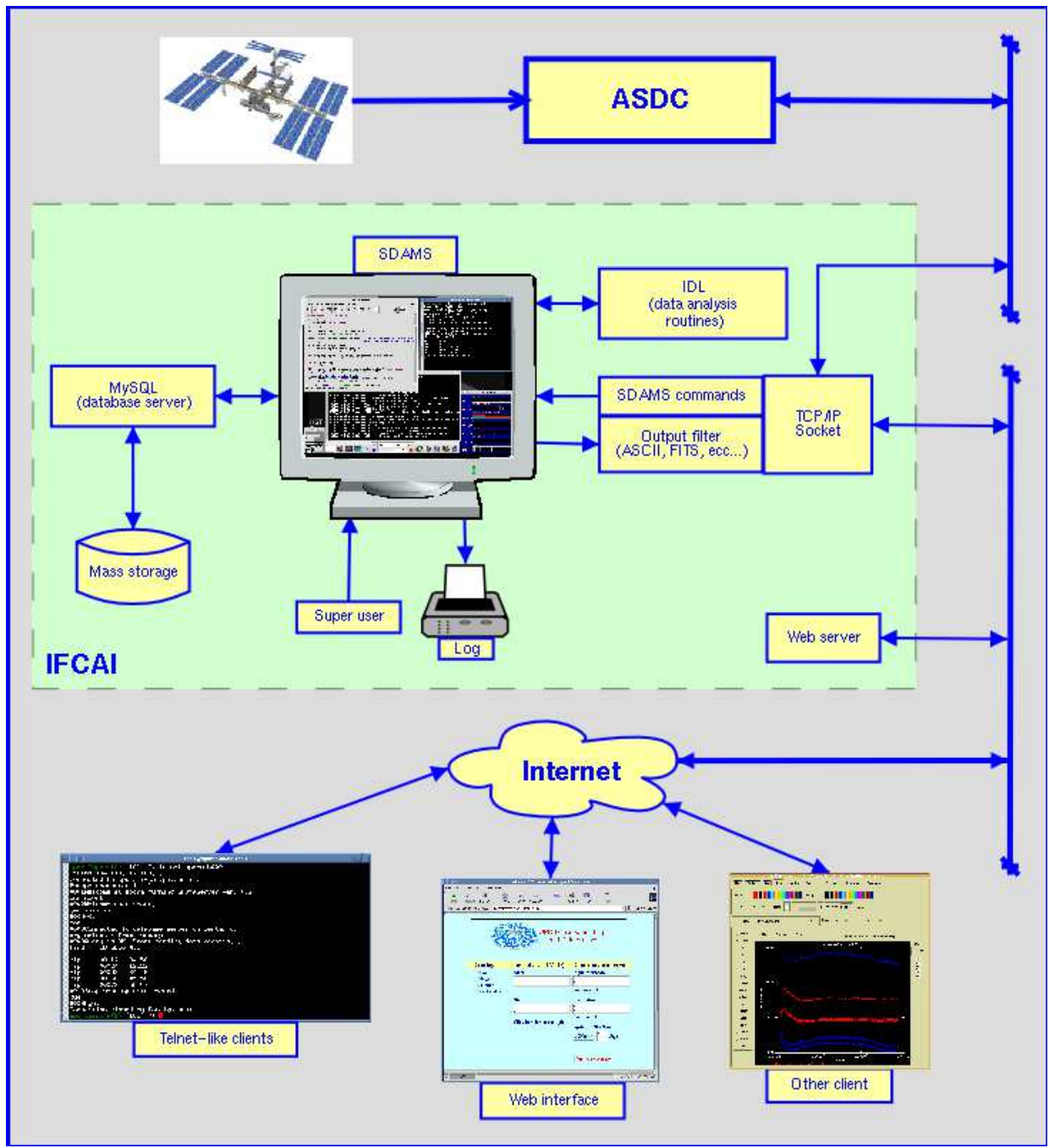}
  \caption{This diagram shows the main components of SDAMS (dashed inset)
 and the way it can interact with external systems (clients).
 The ASDC (ASI Science Data Center) network gives access to the raw data
 downloaded from the International Space Station. Here we assume the
 database resides at IFCAI Institute (Palermo)}
\end{figure}

\begin{figure}
  \includegraphics[height=.5\textheight,keepaspectratio]{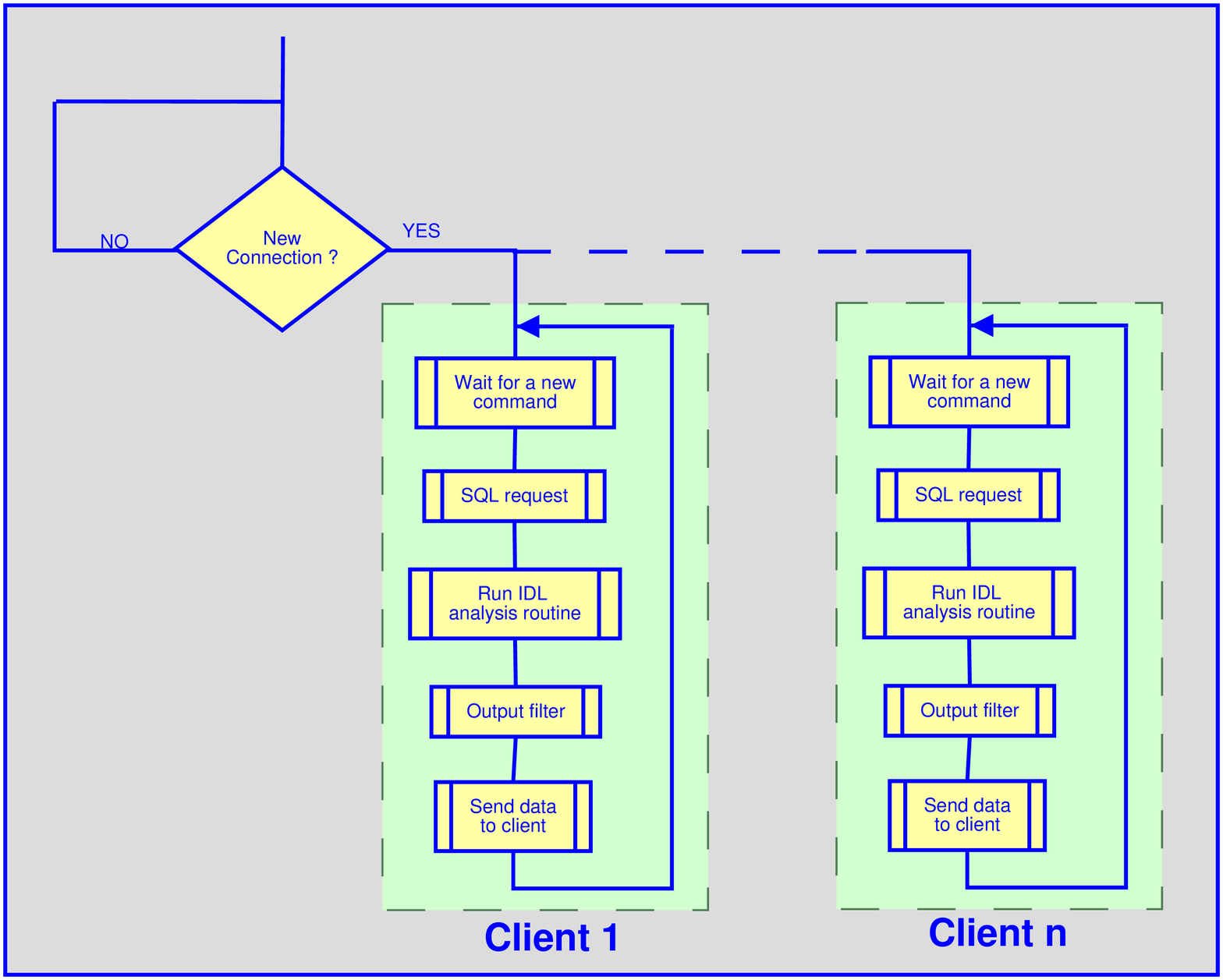}
  \caption{The scheme above shows the parallel running of the threads
 (dashed insets). Each thread is a sub-process (on the server) performing
 the client queries and data output}
\end{figure}

\begin{figure}
  \includegraphics[height=.3\textheight,keepaspectratio]{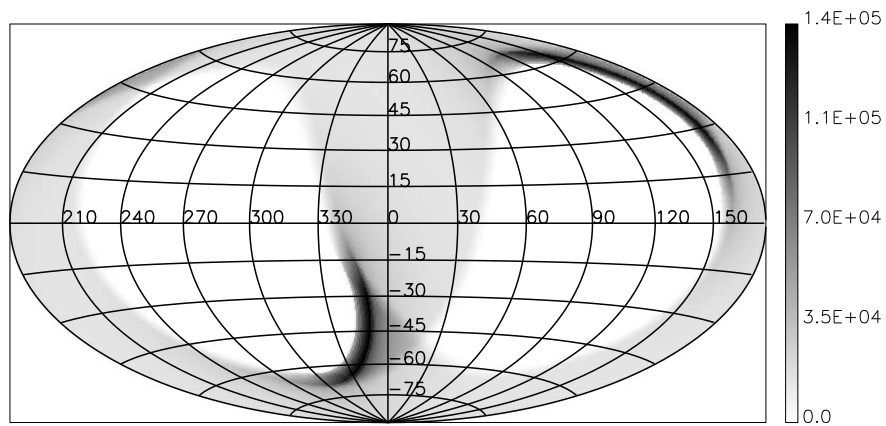}
  \caption{In this image (Galactic coordinates) the SPOrt sky coverage
 after about 20 days of operative life is reported.
 Grey scale is proportional to the integration time
 }
\end{figure}

\begin{theacknowledgments}
SDAMS is part of the SPOrt project and it is supported by
Agenzia Spaziale Italiana (ASI).
\end{theacknowledgments}

% choose bibtex style depending on layout style and options used in
% sample:

\doingARLO[\bibliographystyle{aipproc}]
          {\ifthenelse{\equal{\AIPcitestyleselect}{num}}
             {\bibliographystyle{arlonum}}
             {\bibliographystyle{arlobib}}
          }
\bibliography{sdams}

\hyphenation{Post-Script Sprin-ger}
\begin{thebibliography}{1}
\expandafter\ifx\csname natexlab\endcsname\relax\def\natexlab#1{#1}\fi
\providecommand{\enquote}[1]{``#1''}
\expandafter\ifx\csname url\endcsname\relax
  \def\url#1{\texttt{#1}}\fi
\expandafter\ifx\csname urlprefix\endcsname\relax\def\urlprefix{URL }\fi

\bibitem[Carretti(2001)]{carretti:2002}
Carretti, E., \enquote{The SPOrt project}, in \emph{Astrophysical Polarized
  Backgrounds}, edited by e.~a. S.~Cecchini, AIP Conference Proceedings~0,
  American Institute of Physics, New York, 2001, pp. xx--yy.

\end{thebibliography}

\end{document}